\documentclass{article}
\usepackage[latin1]{inputenc}
\usepackage{graphicx}
\usepackage{chemarr}
\usepackage{float} 
\usepackage{amssymb,amsfonts,amsmath} 
\usepackage{cite} 
\usepackage{hyperref}
\usepackage{url} 
\usepackage{breakurl}

\hypersetup{
    colorlinks,%
    citecolor=blue,%
}

\makeatletter
\renewcommand{\fnum@figure}{\textbf{\figurename\nobreakspace\thefigure}}
\renewcommand{\fnum@table}{\textbf{\tablename\nobreakspace\thetable}}
\makeatother

\title{\textbf{A simple negative interaction in the positive transcriptional feedback of a single gene is sufficient to produce reliable oscillations}}

\author{Jesús M. Miró-Bueno \& Alfonso Rodríguez-Patón \\}
\date{}

\begin{document}

\maketitle
\begin{center}
Departamento de Inteligencia Artificial\\
Facultad de Informática\\
Universidad  Politécnica de Madrid\\
{jmiro@fi.upm.es \& arpaton@fi.upm.es}
\end{center}
\bigskip

\noindent {\small{\textbf{Citation:} Miró-Bueno JM, Rodríguez-Patón A (2011) A Simple Negative Interaction in the Positive Transcriptional Feedback of a Single Gene Is Sufficient to Produce Reliable Oscillations. PLoS ONE 6(11): e27414.

\noindent doi:10.1371/journal.pone.0027414}}\\

\begin{abstract}
\noindent Negative and positive transcriptional feedback loops are present in natural and synthetic genetic oscillators. A single gene with negative transcriptional feedback needs a time delay and sufficiently strong nonlinearity in the transmission of the feedback signal in order to produce biochemical rhythms. A single gene with only positive transcriptional feedback does not produce oscillations. Here, we demonstrate that this single-gene network in conjunction with a simple negative interaction can also easily produce rhythms. We examine a model comprised of two well-differentiated parts. The first is a positive feedback created by a protein that binds to the promoter of its own gene and activates the transcription. The second is a negative interaction in which a repressor molecule prevents this protein from binding to its promoter. A stochastic study shows that the system is robust to noise. A deterministic study identifies that the dynamics of the oscillator are mainly driven by two types of biomolecules: the protein, and the complex formed by the repressor and this protein. The main conclusion of this paper is that a simple and usual negative interaction, such as degradation, sequestration or inhibition, acting on the positive transcriptional feedback of a single gene is a sufficient condition to produce reliable oscillations. One gene is enough and the positive transcriptional feedback signal does not need to activate a second repressor gene. This means that at the genetic level an explicit negative feedback loop is not necessary. The model needs neither cooperative binding reactions nor the formation of protein multimers. Therefore, our findings could help to clarify the design principles of cellular clocks and constitute a new efficient tool for engineering synthetic genetic oscillators.
\end{abstract}


\noindent {\small{\textbf{Abbreviations:} {NTF, negative transcriptional feedback; PTF, positive transcriptional feedback; QSSA, quasi-steady-state assumption}}}

\section*{Introduction}

\begin{figure}[!ht]
\begin{center}
\includegraphics[width=4in]{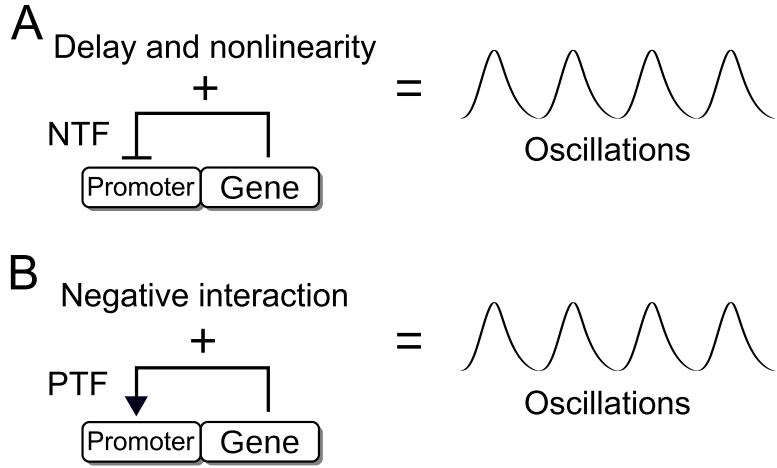}
\end{center}
\caption{\small{{\bf Diagram of one-gene oscillators with negative and positive transcriptional feedbacks.} {\bf A}. Negative transcriptional feedback (NTF) created by a protein that represses the expression of its own gene. This NTF needs time delay and sufficiently strong nonlinearity in the feedback signal transmission in order to produce reliable oscillations. The time delay is created by intermediate reactions, such as the transcription and translation, reversibly phosphorylations or proteins shuttling between the nucleus and the cytoplasm. The nonlinearity can be created by reactions, such as protein cooperativity in the gene repression or formation of protein multimers. {\bf B}. Positive transcriptional feedback (PTF) created by a protein that activates the expression of its own gene. This PTF needs a negative interaction in the feedback signal transmission in order to produce reliables oscillations. The negative interaction can be a degradation, sequestration, or inhibition carried out by a repressor molecule.}}
\label{figure1}
\end{figure}

Cellular clocks control important functions of the cell, such as circadian (24-hour) rhythms, cell cycle, metabolism and signaling. Clock operation appears to involve the coupling of two different types of oscillators. The first are oscillators based on cytoplasmic reactions, such as phosphorylation \cite{Rust} and oxidation \cite {O'Neill_a,O'Neill_b}. The second are genetic oscillators depending on gene expression regulation \cite{Dunlap,Young}. In the last decade several synthetic genetic oscillators have been implemented in the laboratory \cite{Elowitz_a,Atkinson,Fung,Stricker,Tigges,Toettcher,Danino}. The first mathematical model of a genetic oscillator was developed by Goodwin for periodic enzyme production \cite{Goodwin}. This model was the groundwork for subsequent theoretical research on genetic oscillators in living systems, such as fungi and flies \cite{Goldbeter,Ruoff_a,Leloup_a,Ruoff_b,Gonze,Bratsun}. In these models, the rhythms are generated by a gene with a negative transcriptional feedback (NTF) (Fig. \ref{figure1}A). This NTF needs time delay and sufficiently strong nonlinearity in the transmission of the feedback signal for preventing the steady-state stabilization of the system \cite{Griffith_a,Novak}. It has also been analyzed variants, involving two genes, of the model presented in the Fig. \ref{figure1}A \cite{Widder}.

\begin{figure}[!ht]
\begin{center}
\includegraphics[width=3in]{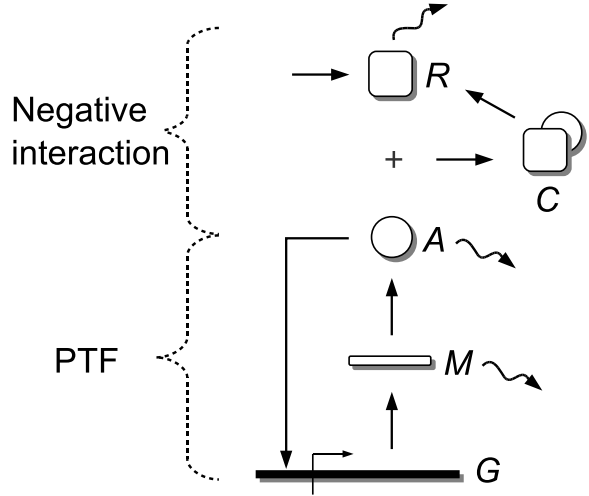}
\end{center}
\caption{\small{{\bf Model of a one-gene oscillator with PTF.} The model is composed of two well-differentiated parts. The first part is a positive feedback loop in which a gene ($G$) is transcribed into mRNA ($M$). In turn, $M$ is translated into protein ($A$). This protein is a transcription factor of its own gene and increases the transcription rate when it binds to the promoter. The positive feedback needs a second part, consisting of a negative interaction in order to obtain reliable oscillations. In this part repressor molecules ($R$) enter the system at a constant rate. $R$ inhibits the function of $A$. Specifically, $R$ binds to $A$ and forms the complex $C$. In this complex, $A$ is not able to bind to its promoter. $R$ is not degraded together with $A$ and can be used several times. Therefore, $R$ can be thought of as a protease, a protein that sequesters $A$ or any other molecule that binds to and inhibits the function of $A$ as explained above. The zigzag arrows stand for degradations. A different version of the model can be formulated with the negative interaction acting over $M$ instead of over $A$.}}
\label{figure2}
\end{figure}

Positive transcriptional feedbacks (PTFs) are also present in many cellular clocks \cite{Reppert, Gallego, Purcell}. Models with two or more genes involving PTFs have been studied in genetic oscillators \cite{Barkai,Smolen,Hasty,Leloup_b,Francois,Guantes,Hong,Conrad,Munteanu}. In these models the PTFs increase the expression of repressor genes. It has been shown how PTFs produce bistability \cite{Becskei,Ferrell}, increase the robustness of cellular clocks \cite{Tsai,Vilar} and could provide robust adaptation to environmental cycles \cite{Mondragon-Palomino}. Previously, it has been demonstrated that a single gene with only PTF does not produce oscillations \cite{Griffith_b}. Here we study a model with a simple condition to produce biochemical rhythms in a single gene with PTF (Fig. \ref{figure1}B).  We chose a circadian period for the oscillator due to its relevance in biological systems. This model is based on two common features of genetic oscillators \cite{Dunlap,Novak,Barkai,Hasty,Vilar}. The first is a PTF created by a protein that activates the transcription of its own gene. The second is a negative interaction in which a repressor inhibits the activity of this protein. We performed stochastic and deterministic simulations that yielded similar results. The stochastic simulations show that the genetic oscillator is robust to noise. This noise is introduced in living cells by the stochasticity of gene expression \cite{Elowitz_b,Swain}. By means of a reduced deterministic model, we show that the oscillations exhibit limit-cycle behavior. This means that if a disturbance is applied to the system, the oscillations return to the original periodic solution \cite{Murray,Strogatz}. Also we show that this biological clock can be classified as a relaxation oscillator \cite{Hasty,Murray,Strogatz}. This type of clock is sometimes called hysteresis oscillator \cite{Barkai,Tyson} or amplified negative feedback oscillator \cite{Novak,Purcell}. The relaxation oscillator comprises fast and slow oscillation creation stages. In our model these oscillations are characterized by sawtooth waveforms. Finally, we explain how the negative interaction works through a comparison with the dynamics of the typical enzymatic reaction. We show that the rate of the negative interaction is amplified by the PTF and has a saturation point.

\section*{Results}

\subsection*{Model and simulations}
The model is a simple one-gene network with two well-differentiated parts (Fig. \ref{figure2}). The first is a PTF created by a protein $A$, which is a transcription factor of its own gene. When this protein binds to its promoter the transcription rate increases. The second part is a negative interaction in which a repressor molecule $R$ prevents $A$ from binding to its promoter. The molecule $R$ can be thought of as a protease, as a protein that sequesters $A$, or as any other molecule that inhibits the function of $A$ as shown in Fig. \ref{figure2}. A different version of the model can be formulated in which the negative interaction acts on the mRNA molecules instead of on protein $A$.

\begin{figure}[!ht]
\begin{center}
\includegraphics[width=6.8in]{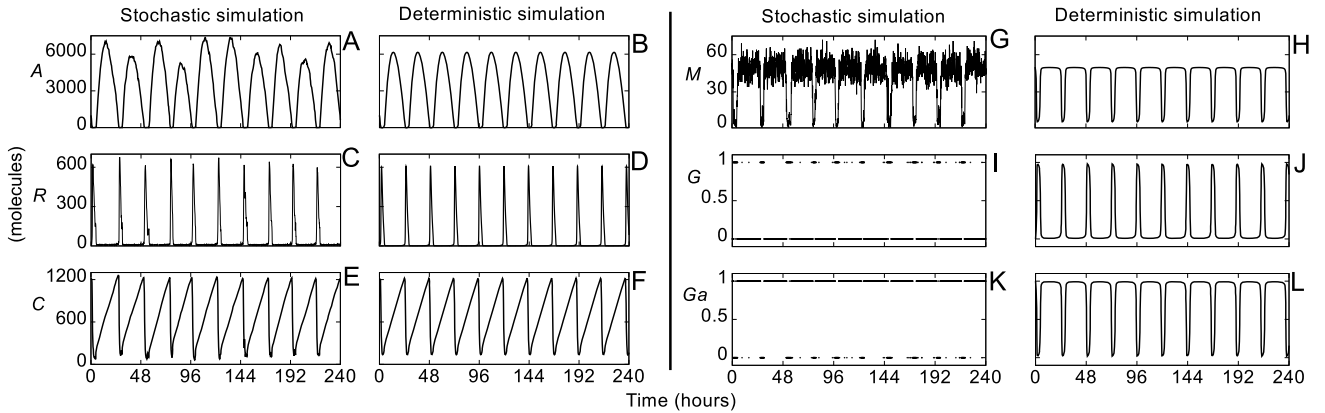}
\end{center}
\caption{\small{{\bf Stochastic and deterministic simulations of the model.} {\bf A, C, E, G, I, K.} Stochastic time evolution of the protein ($A$), repressor ($R$), protein-repressor complex ($C$), mRNA ($M$), gene ($G$) and activated gene ($G_a$), respectively. {\bf B, D, F, H, J, L.} Deterministic time evolution of $A$, $R$, $C$, $M$, $G$, and $G_a$, respectively. In both simulations, the time evolution of $A$ ({\bf A} and {\bf B}), $R$ ({\bf C} and {\bf D}) and $C$ ({\bf E} and {\bf F}) are very similar except for the presence of fluctuations in the stochastic case. This phenomenon is more pronounced in the time evolution of $M$ ({\bf G} and {\bf H}). The oscillations in $C$ show sawtooth waveforms. There is a single gene in the model; hence $\textstyle{G+G_a=1}$ molecule. In the time evolution of $G$ ({\bf I} and {\bf J}) and $G_a$ ({\bf K} and {\bf L}), the stochastic simulation shows discrete transitions between 0 and 1 molecules. By contrast, the deterministic simulation shows unrealistic continuous transitions. The time evolution of $G_a$ shows that the gene is activated most of the time ({\bf K} and {\bf L}).}}
\label{figure3}
\end{figure}

Eleven biochemical reactions provide a full description of the model (see {\eqref{reactions}} in the section \textit{Methods: Biochemical reactions and rates}). The system is assumed to have a uniform mixture of biomolecules. For this reason, we did not take into account diffusion processes. In this approach, the dynamics of the biochemical reactions {\eqref{reactions}} can be described by two different formalisms known as stochastic and deterministic approaches (see {\it Methods: Deteministic and stochastic simulations} for more details). These two approaches can lead to different behaviors. The stochastic dynamics of the reactions {\eqref{reactions}} were simulated using the Gillespie algorithm \cite{Gillespie} and the deterministic dynamics using the following ordinary differential equations:
\begin{equation}
\label{edos}
\begin{aligned}
dG/dt & = -k_1 G A + k_{-1} G_a \\
dG_a/dt & = k_1 G A - k_{-1} G_a \\
dM/dt & = k_2 G + k_3 G_a - k_4 M \\
dA/dt & = -k_1 G A + k_{-1} G_a + k_5 M -k_6 A - k_7 A R\\
dR/dt & = - k_7 A R + k_8 C + k_9 - k_{10} R\\
dC/dt & = k_7 A R - k_8 C,
\end{aligned}
\end{equation}
where the variables and rates are described in the section \textit{Methods: Biochemical reactions and rates}. We used standard values within the diffusion limit for the rates \cite{Gonze,Vilar,Dublanche}. 

The stochastic approach is more realistic than the deterministic simulation because it takes into account the randomness of the chemical reactions. This randomness produces fluctuations in the number of molecules. We fitted the reaction rates to obtain circadian oscillations in the stochastic simulation. Then, we compared the results with the deterministic simulation (Fig. \ref {figure3}). For both simulations the time evolution of the protein ($A$), repressor ($R$), protein-repressor complex ($C$) and mRNA ($M$) are very similar. The main difference is the appearance of fluctuations in the stochastic case around the number of molecules predicted by the deterministic approach. The fluctuations are more evident in the time evolution of $M$ (Fig. \ref {figure3}G) than in the other biomolecules. This is because the number of $M$ molecules oscillates in a lower range than $A$, $R$ and $C$. The oscillations in $C$ are characterized by sawtooth waveforms. On the other hand, there are differences between the stochastic and deterministic time evolution of the gene. There is a single gene in the model, which can be deactivated ($G$) or activated ($G_a$). Therefore, $\textstyle{G+G_a=1}$ molecule. The stochastic simulation shows realistic discrete transitions between 0 and 1 molecules (Figs. \ref{figure3}I and \ref{figure3}K). By contrast the deterministic simulation shows unrealistic continuum transitions (Figs. \ref{figure3}J and \ref{figure3}L). In both cases, however, the qualitative behavior is the same. Most of the time the gene is activated by $A$, although it is deactivated for a short time when the number of $A$ in the oscillations is low. 

\subsection*{Model robustness to noise}

\begin{figure}[H]
\begin{center}
\includegraphics[width=4.5in]{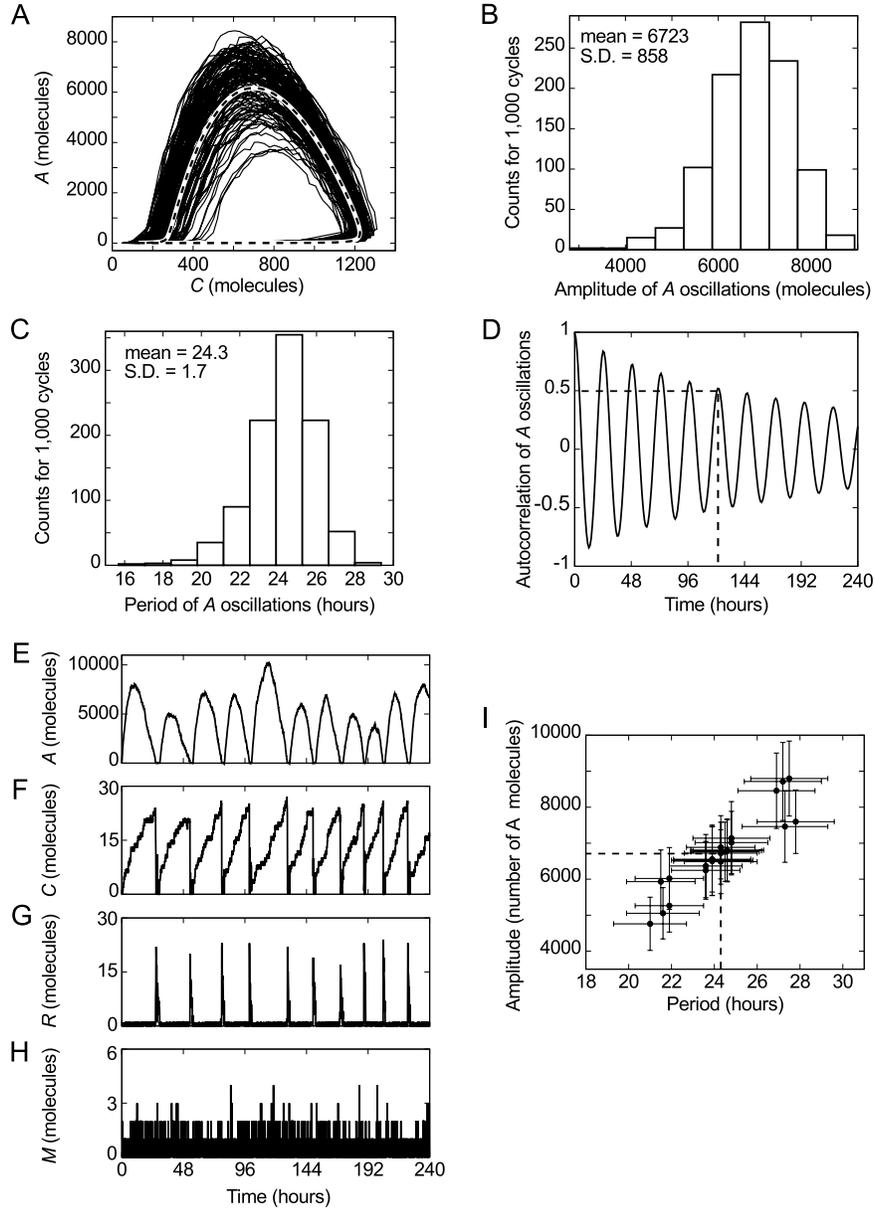}
\end{center}
\caption{\small{{\bf Model robustness to noise.} {\bf A.} Stochastic and deterministic phase plane defined by $C$ and $A$. The general shape of the curves is the same in both cases, but the deterministic curve (dashed line) is well defined because the oscillations are identical. In contrast, the stochastic curve (solid line) spreads around the deterministic one. {\bf B, C.} Amplitude and period histograms of the stochastic simulation of $A$, respectively. {\bf D.} Autocorrelation of the stochastic oscillations in the number of $A$ molecules. The half-life of the autocorrelation is about 120 hours (intersection of dashed lines). {\bf E-H.} Model robustness to intrinsic noise when the number of molecules is low. The changed rates are $\textstyle{k_4=1000}$ hour$^{-1}$, $\textstyle{k_5=5000}$ hour$^{-1}$, $\textstyle{k_7=25.5}$ molecules$^{-1}$ hour$^{-1}$, $\textstyle{k_8=132.6}$ hour$^{-1}$ and $\textstyle{k_9=1}$ molecules hour$^{-1}$. In particular, we multiplied the rates $k_4$ and $k_5$ by 100 to obtain a low number of $M$ molecules. Simultaneously, we multiplied the rates $k_7$ and $k_8$, and divided the rate $k_9$ by 51 to obtain a low number of $R$ and $C$ molecules. The initial conditions are $\textstyle{G_{a0}=1}$ and $\textstyle{G_0=M_0=A_0=R_0=C_0=0}$ molecules. The mean value of $M$ is 0.48 molecules. {\bf I.} Model robustness to extrinsic noise. Scatter plot of amplitude versus period that shows the robustness of the model to parameter variation (data is presented in Table S1). Two stochastic simulations were performed for each parameter in which the value was increased and decreased by 15\%. The x and y coordinates of each data point correspond to the mean values of the period and amplitude, respectively. The horizontal and vertical error bars are the standard deviation of the period and amplitude, respectively. The intersection between dashed lines shows the point obtained without changing the value of any rate (Figs. \ref{figure4}B and \ref{figure4}C). ({\bf B}, {\bf C}, {\bf D} and each data point in {\bf I} were calculated for 1,000 successive cycles. We assumed that a cycle occurs if the number of proteins $A$ increases to 1,000 molecules and then decreases to 700 molecules. The amplitude was calculated as the greatest number of $A$ molecules in each cycle. The period was calculated as the time interval that it takes the number of proteins $A$ to reach 1,000 molecules for the first time in two successive cycles.)}}
\label{figure4}
\end{figure}

The fluctuations in the stochastic simulation are the source of so-called intrinsic noise \cite{Elowitz_b,Swain}. In the genetic oscillator, this intrinsic noise generates variability in both the amplitude and period of the oscillations. The phase plane defined by $C$ and $A$ illustrates this variability very clearly (Fig. \ref{figure4}A). The deterministic phase plane is a well-defined curve because the oscillations are identical (dashed line in Fig. \ref{figure4}A). In contrast, the stochastic phase plane is a curve that spreads around the deterministic curve due to intrinsic noise (solid line in Fig. \ref{figure4}A). We used the amplitude and period histograms, and the autocorrelation function to quantify the effect of this intrinsic noise on $A$ oscillations. The results are similar to circadian models with more chemical reactions \cite{Barkai,Gonze}. The amplitude histogram shows a mean of 6,723 molecules and a standard deviation of 858 molecules (Fig. \ref{figure4}B). The period histogram shows a mean of 24.3 hours and a standard deviation of 1.7 hours (Fig. \ref{figure4}C). In contrast, the absence of intrinsic noise in the deterministic simulation produces identical $A$ oscillations with lower amplitude and period equal to 6,164 molecules and 23.6 hours, respectively. On the other hand, the autocorrelation function shows a half-life time of about 120 hours (Fig. \ref{figure4}D). 

The stochastic approach produces good oscillations in $A$ even when there are fewer than 30 molecules of M, R and C. (Figs. \ref{figure4}E-H). We changed the value of some rates to obtain this simulation as in ref. \citen{Vilar} (see caption of Fig. \ref{figure4}). In the deterministic approach, where intrinsic noise is not present, these changes do not alter the dynamics of $A$ significantly and produce a low number of $M$, $R$, and $C$ molecules. In particular, the amplitude and the period are slightly lower (Fig. S1). In the stochastic simulation the rate changes reduce the amplitude and period means to 6,166 molecules and 21.3 hours, respectively (Fig. S2). The effects of intrinsic noise is now more pronounced because the number of $M$, $R$, and $C$ molecules is low. This is reflected in an increase of the amplitude and period standard deviations to 2,132 molecules and 5.2 hours, respectively (Fig. S2). 

In cells, there are also fluctuations in the number (or activity) of molecules such as polymerases, ribosomes and degradation machinery. These fluctuations are the source of so-called extrinsic noise \cite{Elowitz_b,Swain}. We performed stochastic simulations varying the parameters in order to account for some aspect of extrinsic noise in the robustness study of the model. The results show that this oscillator is robust to small parameter variations (Fig. \ref{figure4}I) like more other complex models of genetic oscillators \cite{Smolen}. The largest amplitude and period changes occurred for variations in $k_3$ (see Table S1). The changes in the mean period and amplitude were always less than 15\% and 31\%, respectively. Particularly, variations in the rates $k_1$, $k_{-1}$, $k_2$, $k_6$, $k_7$ and $k_{10}$ produced changes of less than 3\% and 8\% in the mean period and amplitude, respectively. The changes in the standard deviation of the period and the amplitude were always less than 13\% and 27\%, respectively.

\subsection*{Reduced deterministic model}
\begin{figure}[!ht]
\begin{center}
\includegraphics[width=6in]{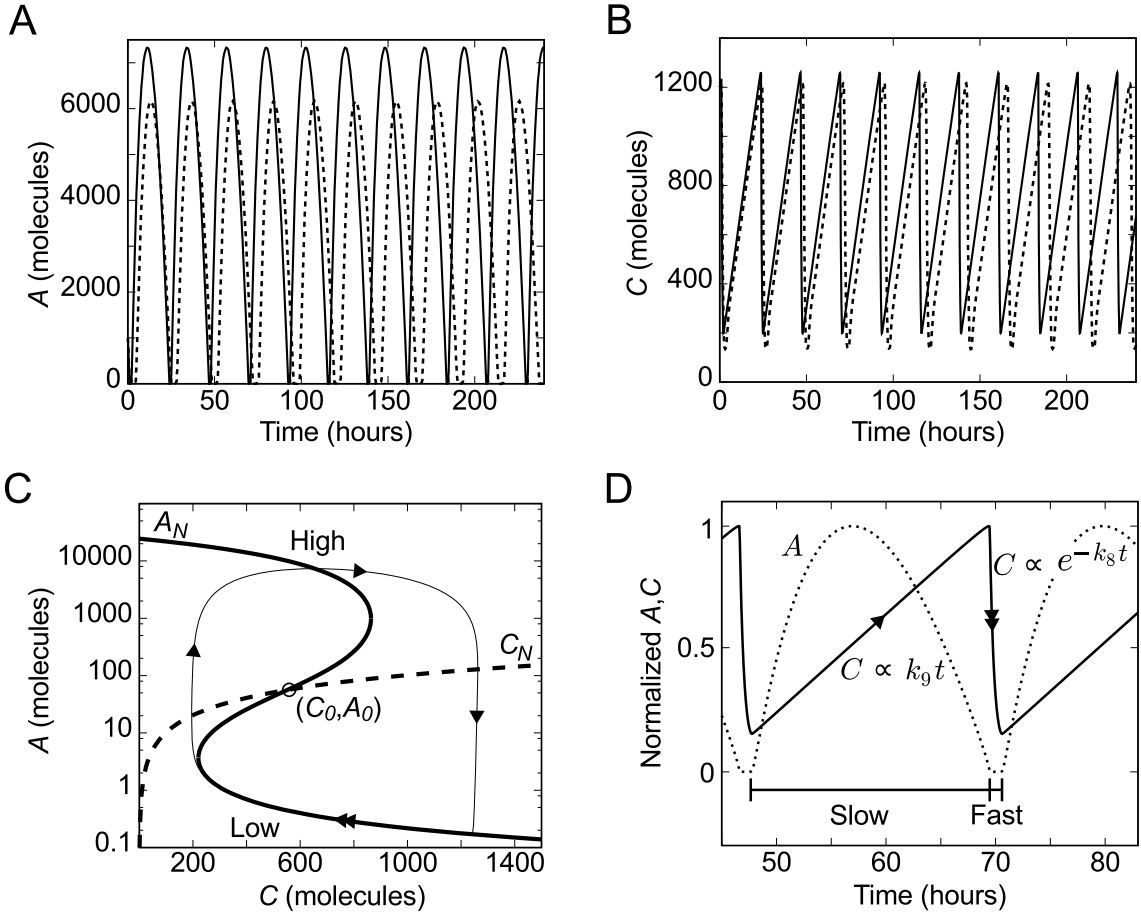}
\end{center}
\caption{\small{{\bf Model of the genetic oscillator reduced by QSSA.} {\bf A, B.} Comparison between the reduced (solid line) and complete (dashed line) deterministic simulation of the time evolution of $A$ and $C$, respectively. {\bf C.} Phase plane. Limit cycle (thin solid line) and nullclines $A_N$ (thick solid lines) and $C_N$ (thick dashed line). The unstable fixed point of the system (marked by circle $\circ$) is $C_0=552.4$ and $A_0=56.3$ molecules. The nullclines $A_N$ and $C_N$ are the solution of equations $dA/dt=0$ and $dC/dt=0$, respectively. The two branches in the nullcline $A_N$ are called ``high'' and ``low''. {\bf D.} Slow and fast stages in the reduced system. The solid line is $C$ and the dotted line is $A$. $C$ exhibits a sawtooth waveform. (The arrows in {\bf C} and {\bf D} represent the direction of the oscillations. One and two arrows mean slow and fast stages, respectively).}}
\label{figure5}
\end{figure}

To identify the types of biomolecules mainly responsible for oscillations, it is useful to reduce the deterministic model by means of the quasi-steady-state assumption (QSSA) \cite{Murray,Fall}. This approximation differentiates between fast and slow variables. The greater the time-scale separation between the variables the more accurate the approximation is. In this approach it is assumed that fast variables quickly reach the equilibrium, i.e., their derivatives are zero. This assumption means that slow variables are responsible for the system dynamics. In this model, we assumed that the fast variables are $G$, $G_a$, $M$ and $R$, and the slow variables are $A$ and $C$. Then, the set of Eq. \eqref{edos} can be simplified to
\begin{equation}
\label{edos_reduced}
\begin{aligned}
dA/dt & = \frac{\alpha + \beta A}{\gamma + A} - A \frac{k_9 + k_8 C}{\delta + A} - k_6 A\\
dC/dt & = \frac{k_9A - \delta k_8 C}{\delta + A},
\end{aligned}
\end{equation}
where $\textstyle{\alpha=G_tk_{-1}k_2k_5/k_1k_4}$, $\textstyle{\beta=G_tk_3k_5/k_4}$, $\textstyle{\gamma=k_{-1}/k_1}$, $\textstyle{\delta=k_{10}/k_7}$ and $\textstyle{G_t=G+G_a}$. A good way to check if this approximation is correct is to compare the numerical solution of the complete and the reduced systems. Both numerical solutions agree except for quantitative differences in the period and the amplitude (Figs. \ref{figure5}A and \ref{figure5}B). These differences are due to the fact that the time-scale separation between fast and slow variables is not large enough for QSSA to be more accurate. Despite these differences, we can conclude that $A$ and $C$ are mainly responsible for the system dynamics. The other types of biomolecules can be considered to be at equilibrium. The fluctuations in the fast variables do not significantly affect the system dynamics \cite{Vilar}. This explains the robustness of the model when the number of molecules is low (Figs. \ref{figure4}E-H). In fact, the system produces reliable oscillations even if the average of $M$ is less than one molecule (Fig. \ref{figure4}H), and, surprisingly, even when the driven $C$ molecules oscillate in a range of less than $30$ molecules (Fig. \ref{figure4}F).

The oscillations in the reduced deterministic model exhibit limit-cycle behavior (thin solid line in Fig. \ref{figure5}C). Therefore, if an external disturbance is applied to the oscillator, the system will go back to oscillating with the period and amplitude of its limit cycle. The unstable fixed point of the system is $C_0=552.4$ and $A_0=56.3$ molecules (circle in Fig. \ref{figure5}C). For a bifurcation analysis of parameters $k_8$ and $k_9$ indicating the range of values that produces limit-cycle oscillations, see \textit{Methods: Bifurcation diagram}.

This genetic clock belongs to the so-called relaxation oscillators \cite{Hasty,Murray,Strogatz}. The mechanism responsible for the oscillations is represented by the nullclines $A_N$ and $C_N$ (Fig. \ref{figure5}C). These nullclines are the solution of the equations $\textstyle{dA/dt=0}$ and $\textstyle{dC/dt=0}$, respectively. The nullcline $C_N$ is a straight line and the nullcline $A_N$ has the characteristic ``Z'' shape of relaxation oscillators \cite{Tyson,Strogatz,Murray}. The shape of the $A$ nullcline is the same as the hysteresis diagram obtained if $C$ is assumed constant (Fig. S6). Therefore, this genetic clock contains some features of hysteresis in its oscillatory mechanism. The $A$ nullcline has two branches that we can call ``high'' and ``low'' (Fig. \ref{figure5}C). These branches are steady states if the $C$ is a constant (Fig. S6). In each oscillation the system switches from one branch to the other using the number of $C$ molecules as a transient signal. This process can be explained following the limit-cycle trajectory. When $A$ and $C$ are about 1 and 200 molecules, respectively, their number increases until $A$ reaches its maximum of about 7,330 molecules and $C$ reaches about 650 molecules. This is the transient from the low to the high branch. Then, the number of $A$ molecules is reduced to about 0 molecules, whereas $C$ reaches its maximum of about 1,260 molecules. This is the transient from the high to the low branch. Finally, the number of $C$ molecules is quickly reduced and the trajectory moves along the nullcline $A_N$, returning to the starting point where a new cycle begins. 

This genetic clock is characterized by containing fast and slow stages. The time evolution of $C$ shows these two well-differentiated stages (Fig. \ref{figure5}D). In the slow stage $\textstyle{A \gg \delta}$ and $\textstyle{k_9 A \gg \delta k_8 C}$, then the second differential equation in \eqref{edos_reduced} can be approximated by $\textstyle{dC/dt \approx k_9}$. In this stage, therefore, the number of $C$ molecules increases linearly according to equation $\textstyle{C \propto k_9 t}$. In the fast stage $\textstyle{A \ll \delta}$ and $\textstyle{k_9 A \ll \delta k_8 C}$, then the second differential equation in \eqref{edos_reduced} can be approximated by $\textstyle{dC/dt \approx -k_8C}$. In this stage, the number of $C$ molecules decays exponentially according to equation $\textstyle{C \propto \exp(-k_8 t)}$. The two stages play different roles. The slow stage is characterized by the formation of a pulse of $A$ molecules. On the other hand, the decay of $C$ into $R$ in the fast stage provides the necessary conditions for a new pulse. These two stages produce oscillations in $C$ with sawtooth waveforms (solid line Fig. \ref{figure5}D).

\subsection*{How the negative interaction works}
The negative interaction decreases the number of free $A$ molecules and takes the system back to the start of a new cycle. The detailed explanation of how this interaction works is related to the dynamics of the typical enzymatic reaction $\textstyle{S + E \xrightleftharpoons[c_{-1}]{c_1} D \xrightarrow{c_2} P + E}$, where $S$, $E$, $D$ and $P$ are the substrate, enzyme, complex substrate-enzyme and product, respectively. The total number of enzymes ($\textstyle{E_t=E+D}$) is constant in the system. The rate of catalysis in this reaction is defined as $\textstyle{v\equiv dP/dt = c_2 D}$. The value of this rate can be approximated by QSSA. The result of this approximation is the well-known Michaelis-Menten equation $\textstyle{v \approx V_{max}S/(K_M + S)}$, where $\textstyle{V_{max}=c_2E_t}$ and $\textstyle{K_M= (c_{-1}+c_2)/c_1}$ \cite{Murray}. In this equation, the rate $v$ increases asymptotically as a function of $S$. The rate $v$ reaches a maximum value ($V_{max}$) when the amount of $S$ is large compared with the constant $K_M$. In this situation, the enzymes are saturated because most are part of complex $D$, and adding more $S$ does not increase the rate $v$. Therefore, $\textstyle{D\approx E_t}$, and the rate of the catalysis $v$ reaches the constant value $c_2 E_t$.

\begin{figure}[!ht]
\begin{center}
\includegraphics[width=6in]{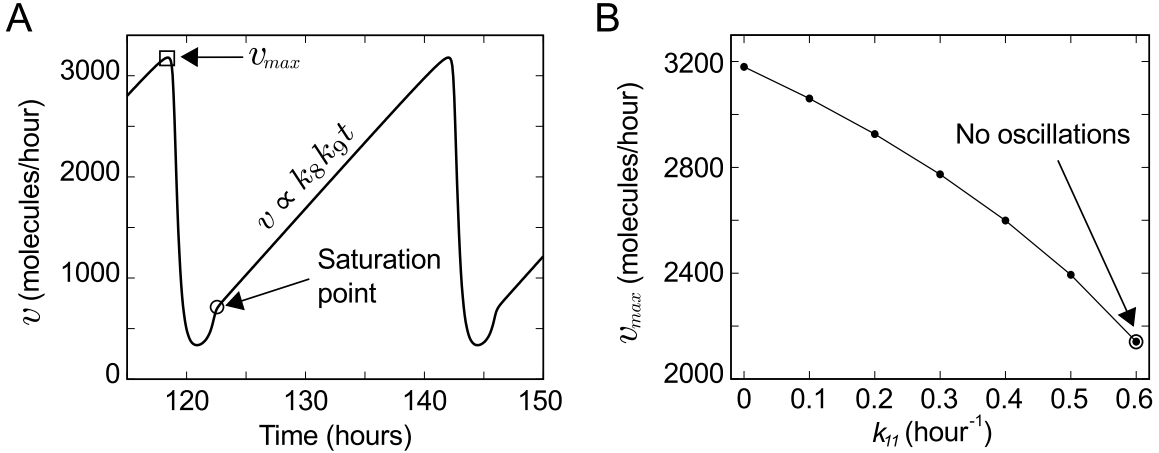}
\end{center}
\caption{\small{{\bf Rate of the negative interaction.} {\bf A.} Rate of the negative interaction ($v = k_8 C$). This rate represents the number of degraded $A$ molecules per hour. The graph was plotted by multiplying the number of $C$ molecules in Fig. 3F by $k_8$. The circle ($\circ$) indicates the saturation point. At the saturation point the rate increases linearly ($\textstyle{v \propto k_8 k_9 t}$) because new $R$ molecules enter the system at rate $k_9$. The square ($\square$) indicates the maximum rate of the negative interaction ($v_{max}=$ 3,180 molecules/hour). {\bf B.} Plot of $v_{max}$ against $k_{11}$, where $k_{11}$ is the rate of the reaction: $\textstyle{C \xrightarrow{} \phi}$. Each point corresponds to a deterministic simulation with $\textstyle{k_{11}}$ equal to 0.0, 0.1, 0.2, 0.3, 0.4, 0.5 and 0.6 hour$^{-1}$, respectively (see Fig. S5 for more detailed information). The oscillations stop when $\textstyle{k_{11}}=$ 0.6 hour$^{-1}$ ($\circ$) (Fig. S5G).}}
\label{figure6}
\end{figure}

In our model, the negative interaction is $\textstyle{A + R \xrightleftharpoons[k_{-7}]{k_7} C \xrightarrow{k_8} R}$, where we assumed $\textstyle{k_{-7}=0}$ to simplify the model. We can think of $A$, $R$, and $C$ as $S$, $E$ and $D$, respectively. Therefore, the rate of the negative interaction can be defined as $v \equiv k_8 C$ (Fig. \ref{figure6}A). This rate represents the number of degraded $A$ molecules per hour. The negative interaction works as follows. The number of $A$ molecules increases quickly due to the positive feedback. This rise causes most of the $R$ molecules to bind to $A$ molecules forming the complex $C$. At this point, the system reaches the saturation level (circle in Fig. \ref{figure6}A). The total number of repressor molecules in the system is $R_t=R+C$. Therefore, at the saturation point, $\textstyle{C \approx R_t}$ and the rate $v$ reaches the value ${k_8 R_t}$. The negative interaction is not fast enough to decrease the growth of $A$ molecules immediately after the saturation point is reached. This is because the number of $R_t$ molecules is low at this point. Nevertheless, new $R$ molecules enter the system at rate $k_9$. Therefore, $R_t$ increases linearly over time ($\textstyle{R_t \propto k_9 t}$) compared with the enzymatic reaction in which $E_t$ is constant. This means that the rate of the negative interaction increases linearly according to equation $\textstyle{v \propto k_8 k_9 t}$.  The value of $v$ increases until the negative interaction is fast enough to reduce the number of $A$ molecules and take the system back to the start of a new cycle. The maximum rate reached by the negative interaction is $v_{max}=$ 3,180 molecules/hour (square in Fig. \ref{figure6}A). 

In this model there is not an explicit negative feedback loop at the genetic level. It has been conjectured that all biochemical oscillators involve some sort of negative feedback loop \cite{Novak}. In this genetic clock, an effective negative feedback loop appears in the reduced model (see the section \textit{Methods: The Jacobian matrix}). Intuitively, this effective negative feedback loop can be explained as follows: when $C$ is rare, $A$ is increased by the positive feedback. This rise in the production of $A$ leads to the accumulation of $C$, which in turn increases $v$. This accumulation of $C$ increases until the negative interaction is fast enough to reduce the number of $A$ molecules. In this model, we assumed that $C$ is not degraded. If this complex is degraded according to the reaction $\textstyle{C \xrightarrow{k_{11}} \phi}$, $v$ increases at a slower rate, and its maximum value ($v_{max}$) is reduced (Figs. \ref{figure6}B and S5). The oscillations stop when $\textstyle{k_{11}=0.6}$ hour$^{-1}$ (Fig. S5G), because not enough $C$ is accumulated in order to increase $v$.

This genetic oscillator does not need cooperative binding reactions nor the formation of protein multimers, in contrast to the one-gene oscillator with TNF (Fig. \ref{figure1}A). It has been demonstrated that protein sequestration produces an effective high nonlinearity \mbox{\cite{Buchler_a,Buchler_b}}. But this high nonlinearity is not observed if the repressor molecule is recycled \cite{Buchler_a}. In our model the repressor $R$ can be used several times. Therefore, the negative interaction does not produce an effective high nonlinearity (see \textit{Supporting Information}: Text S1).

\section*{Discussion}
Genetic networks with NTFs and PTFs play an important role in cellular clocks. In this paper, we provided a simple model illustrating that a single gene with PTF has also the potential to produce reliable oscillations. The sufficient additional requirement is a simple and usual negative interaction of degradation, sequestration or inhibition acting on the positive feedback signal. The model presented in this article has a different oscillatory mechanism than the well-established NTF one-gene oscillator model. Our model can be classified as a relaxation oscillator. A two-gene model has been proposed as a different way of producing reliable circadian oscillations in cellular clocks \cite{Barkai}, which also is a relaxation oscillator. This two-gene model is important because it is robust to noise \cite{Vilar}. The model introduced in this paper is a simpler way to produce relaxation oscillations than the previous two-gene oscillator. A comparison with our model reveals that the activation of the repressor gene is not a necessary condition to produce reliable circadian oscillations in the two-gene oscillator. We demonstrated that our model produces circadian oscillations that are just as robust to noise as the two-gene oscillator and other more complex models \cite{Smolen,Gonze}. Similarly to the two-gene oscillator, our model produces good oscillations when the average number of mRNA molecules is less than one. In fact, the number of proteins oscillates satisfactorily even when the other types of molecules involved in the clock are less than 30. Therefore, this model is a simpler genetic relaxation oscillator than the current two-gene clocks \cite{Purcell}. Our model does not need the activation of a second repressor gene by the PTF, cooperative binding reactions nor the formation of protein multimers.

A single gene with PTF and a negative interaction in the feedback signal is an alternative and simple way of generating reliable oscillations. Our study suggests that PTF, besides increasing robustness in cellular clocks, could be more directly and deeply involved in the production of oscillations than at first thought. Further research is necessary to elucidate the presence and the role of this genetic oscillator in natural cellular clocks. On the other hand, thanks to its simplicity, this model has the potential to be a new tool for engineering synthetic genetic oscillators. In this case the period and amplitude of the oscillations could be possibly controlled by externally manipulating the entry rate of the repressor molecules.

\section*{Methods}
\subsection*{Biochemical reactions and rates}
The biochemical reactions that fully describe the model in the Fig. \ref{figure2} are as follows:
\begin{equation}
\label{reactions}
\begin{alignedat}{2}
&\text{Activation/deactivation: } &\quad &G + A \xrightleftharpoons[k_{-1}]{k_1} G_a \\
&\text{Slow transcription: } && G \xrightarrow{k_2}  G + M \\
&\text{Fast transcription: } && G_a \xrightarrow{k_3}  G_a + M \\
&\text{mRNA degradation: } && M \xrightarrow{k_4} \phi \\
&\text{Translation: } && M \xrightarrow{k_5} M + A \\
&\text{$A$ degradation: } && A \xrightarrow{k_6} \phi \\
&\text{Complex creation: } && R + A \xrightarrow{k_7} C \\
&\text{Complex decay into $R$: } && C \xrightarrow{k_8} R \\
&\text{$R$ creation (or entry): } && \phi \xrightarrow{k_9} R \\
&\text{$R$ degradation (or exit): } && R \xrightarrow{k_{10}} \phi,
\end{alignedat}
\end{equation}
where $G$ denotes the gene without $A$ bound to its promoter, $M$ denotes mRNA transcribed from $G$, $A$ denotes the activator protein translated from $M$, $G_a$ denotes the gene with $A$ bound to its promoter, $R$ denotes the repressor and $C$ denotes $R$ bound to $A$. All the biochemical species are measured in molecules. The description of the rates is as follows: $k_1$ is the binding rate of $A$ to the promoter of $G$, $k_{-1}$ is the unbinding rate of $A$ from the promoter of $G$, $k_2$ is the basal transcription rate, $k_3$ is the activated transcription rate, $k_4$ is the degradation rate of $M$, $k_5$ is the translation rate, $k_6$ is the degradation rate of $A$, $k_7$ is the binding rate of $R$ to $A$, $k_8$ is the decay rate of $C$ into $R$, $k_9$ is the creation (or entry) rate of $R$ and $k_{10}$ is the degradation (or exit) rate of $R$.

We used standard values within the diffusion limit for the rates \cite{Gonze,Vilar,Dublanche}. They are as follows: $\textstyle{k_1=1}$ molecules$^{-1}$ hour$^{-1}$, $\textstyle{k_{-1}=50}$ hour$^{-1}$, $\textstyle{k_2=50}$ hour$^{-1}$, $\textstyle{k_3=500}$ hour$^{-1}$, $\textstyle{k_4=10}$ hour$^{-1}$, $\textstyle{k_5=50}$ hour$^{-1}$, $\textstyle{k_6=0.1}$ hour$^{-1}$, $\textstyle{k_7=0.5}$ molecules$^{-1}$ hour$^{-1}$, $\textstyle{k_8=2.6}$ hour$^{-1}$, $\textstyle{k_9=51}$ molecules hour$^{-1}$ and $\textstyle{k_{10}=1}$ hour$^{-1}$. The cell has a single copy of the gene: $\textstyle{G_t=G+G_a=1}$ molecule. The initial conditions are: $\textstyle{G_0=0}$, $\textstyle{G_{a0}=1}$, $\textstyle{M_0=5}$, $\textstyle{A_0=1000}$, $\textstyle{R_0=5}$, and $\textstyle{C_0=1200}$ molecules. The initial conditions have been chosen to obtain a first cycle with an amplitude similar to the limit-cycle oscillations. Note that the rates $k_1$ and $k_7$ include the volume of the system $V$. Hence, these rates can be written as $\textstyle{k_1=k^*_1/V}$ and $\textstyle{k_7=k^*_7/V}$, where the rates $\textstyle{k^*_1}$ and $\textstyle{k^*_7}$ are expressed in M$^{-1}$ hour$^{-1}$. In order to generate circadian oscillations, first, we varied all the reaction rates, according to the values used in refs. \citen{Gonze}, \citen{Vilar} and \citen{Dublanche}, until we got oscillations with a period of around 24 hours in the stochastic simulation. Then we fine-tuned the oscillations varying rates $k_8$ and $k_9$ until a period closer to 24 hours was achieved.

\subsection*{Deteministic and stochastic simulations}
Models based on chemical reactions in a well stirred system are usually described by two different formalisms from a mathematical point of view:

Deterministic: this formalism is suitable for large numbers of molecules. It is described by a set of coupled ordinary differential equations that follow the law of mass action. These equations are called \textit{reaction rate equations} and they can only be solved analytically for simple systems. For more complex systems numerical methods are necessary. In this approach the amount of each chemical species and the time are continuous. The velocity at which reactions occur is given by the reaction rate constants $k$, or simply \textit{rate}.

Stochastic: this formalism is suitable for small numbers of molecules because it takes into account the randomness of the chemical reactions. It is described by the so-called \textit{master equation}, which is the time evolution of the probability that the system has a certain number of molecules of each chemical species at time $t$. Few systems can be solved analytically with the master equation. It is possible, however, to simulate the stochastic behaviour with the Gillespie algorithm \cite{Gillespie}. In this approach the amount of each chemical species and the time are discrete, and the rates $k$ turn into probabilities.

\subsection*{Bifurcation diagram}
We calculated the bifurcation diagram for parameters $k_8$ and $k_9$. These are key parameters for two reasons. First, the rate of the negative interaction $v$ is proportional to $k_8$ and $k_9$ when the saturation point is reached. Second, the fast and slow stages in the relaxation oscillations depend on $k_8$ and $k_9$, respectively. Specifically, we studied the range values of $k_9$ that produce stable oscillations through a bifurcation diagram. Then we studied how this range changes when the parameter $k_8$ varies.

The bifurcation diagram of the reduced model depending on $k_9$ shows two Hopf bifurcation points (Fig. S3A). The first Hopf bifurcation appears at $\textstyle{k_9=4.78}$ molecules hour$^{-1}$ and the second at $\textstyle{k_9=217.6}$ molecules hour$^{-1}$. Most of the values of $k_9$ between these two points produce stable oscillations. Only for a short range of values around these points are the oscillations unstable (white circles in Fig. S3A). The oscillations have an amplitude of from 2,000 to 16,000 molecules, and a period of from 7 to 170 hours (Fig. S3B). The velocity of the reaction $\textstyle{\phi \xrightarrow{k_9} R}$ in \eqref{reactions} does not depend on any biomolecule involved in the oscillator. Therefore, parameter $k_9$ can be interpreted as an external signal controlling the behaviour of the clock.

The variation of parameter $k_8$ changes the position of the two Hopf bifurcation points (white circles in Fig. S4). The different positions of these points define the regions with stable oscillations depending on the values of $k_8$ and $k_9$ (regions I and II in Fig. S4). If parameter $k_8$ is increased, the range of values of $k_9$ that produces stable oscillations decreases. This range shrinks faster if $k_8$ is greater than 20 hour$^{-1}$. We plotted an equivalent graph for the stochastic model because it is more realistic than the reduced graph (black circles in Fig. S4). In particular, we assumed that oscillations occurs in a region if the correlation in the first period is greater than 0.2. The stochastic model produces oscillations in the regions II and III (Fig. S4). The range of oscillations in the complete deterministic model is close to the region II.

\subsection*{The Jacobian matrix}

The Jacobian matrix of the reduced system \eqref{edos_reduced} is:
\begin{equation}
\label{jacobian}
J=
\begin{pmatrix}
a_{11} & a_{12} \\
a_{21} & a_{22}
\end{pmatrix}
=
\begin{pmatrix}
\displaystyle{\frac{\gamma\beta - \alpha}{(\gamma + A)^2} - \frac{\delta(k_9+k_8C)}{(\delta+A)^2} - k_6} &  \displaystyle{-\frac{k_8A}{\delta+A}} \\
 \displaystyle{\frac{\delta(k_9+k_8C)}{(\delta+A)^2}} &  \displaystyle{-\frac{\delta k_8}{\delta+A}}
\end{pmatrix}
,
\end{equation}
where the element $a_{12}$ and $a_{22}$ are always negative, the element $a_{21}$ is always positive and the element $a_{11}$ can be positive or negative depending on the values of the rates. With the rates given in the section \textit{Methods: Biochemical reactions and rates} and the fixed point of the reduced system (Fig. \ref{figure5}C) the sign pattern for the Jacobian matrix is:
\begin{equation}
\label{jacobian_sign}
J=
\begin{pmatrix}
+ & - \\
+ & -
\end{pmatrix}
.
\end{equation}
A two-component negative feedback loop is created in the reduced model because $a_{12}a_{21}<0$ (see Chapter 9 of the reference \cite{Fall}). The Jacobian matrix \eqref{jacobian_sign} has a tipically sign pattern that produces Hopf bifurcation in chemical systems with two variables {\cite{Fall,Murray}}. The two-component systems with this sign pattern in the Jacobian matrix are called activator-inhibitor models \cite{Fall}.

\subsection*{Software}
Code for stochastic and deterministic simulations was written in FORTRAN and XPPAUT (\url{http://www.math.pitt.edu/~bard/xpp/xpp.html}), respectively. Simulations have been contrasted using CAIN software (\url{http://cain.sourceforge.net/}). The stability analysis to determine steady states and limit cycles was performed with XPPAUT. The histograms and autocorrelation function were plotted using FORTRAN and GNU Octave (\url{http://www.gnu.org/software/octave/}). The code for complete and reduced deterministic simulations in XPPAUT is available in File S1 and File S2. The code for stochastic and deterministic simulations in CAIN is available in File S3.

\section*{Funding}
This research has been partially funded by Spanish Ministry of Science and Innovation (MICINN) grant BES-2007-16220 and project TIN2009-14421, by the UPM and Madrid Regional Government and by research project BACTOCOM funded by European Commission under FP7, FET proactive program.

\section*{Acknowledgments}
Part of this work was carried out by JMMB during his stay at the Novel Computation Group led by Dr. Martyn Amos at MMU. We would also like to thank Rachel Elliott and Niall Murphy for polishing the English, and an anonymous reviewer for useful comments on the manuscript.

\bibliographystyle{plos2009}
\bibliography{thereferences}

\newpage 

\section*{Supporting Information: Figures}

\begin{center}
\includegraphics[width=3in]{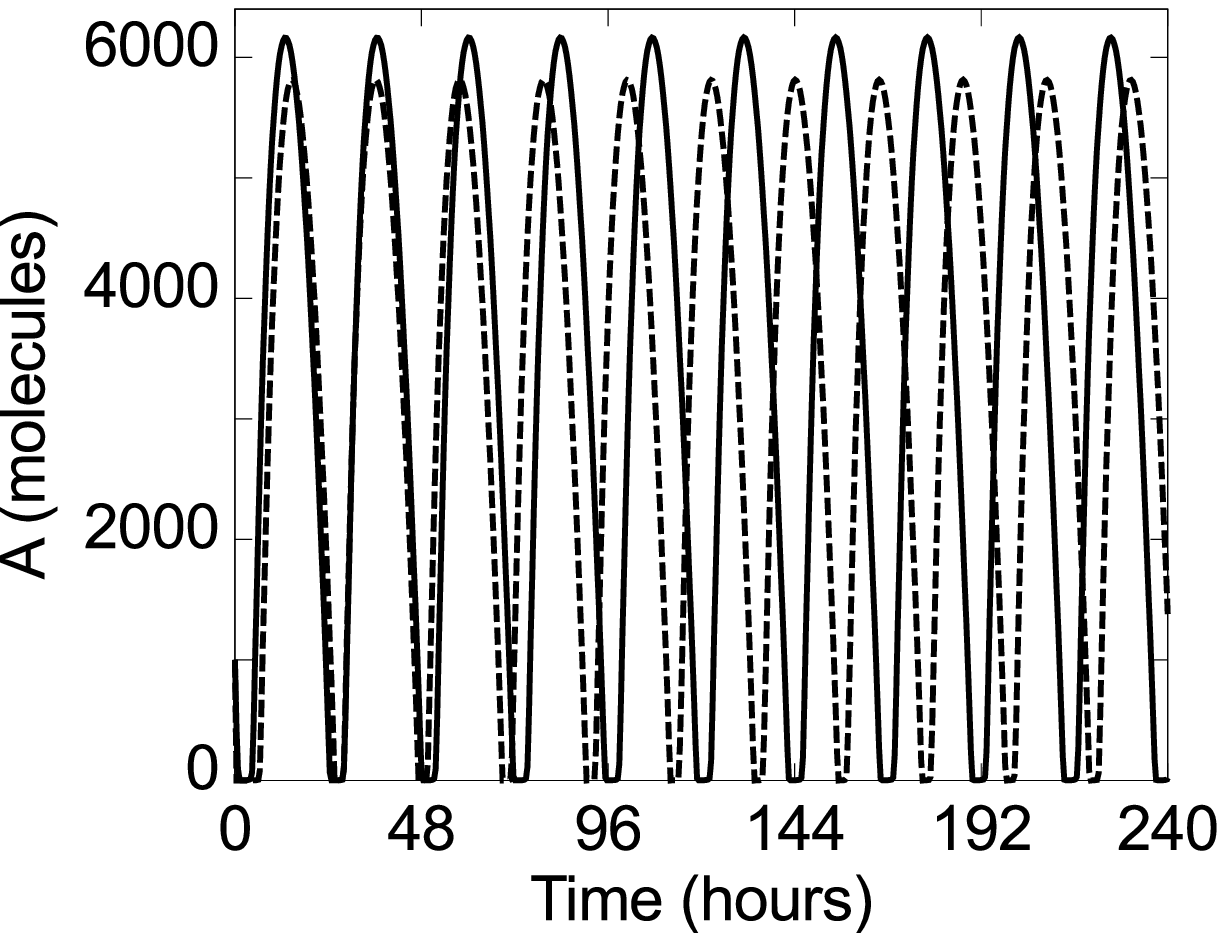}
\end{center}

\noindent {\bf Figure S1. Time evolution of $\boldsymbol{A}$ with and without a low number molecules.} Comparison between deterministic simulation of the time evolution of $A$ with (dashed line) and without (solid line) a low number of $M$, $R$, and $C$ molecules. (Solid line graph: the values of the parameters are as in the section \textit{Methods: Biochemical reactions and rates}. Dashed line graph: the changed rates are $\textstyle{k_4=1000}$ hour$^{-1}$, $\textstyle{k_5=5000}$ hour$^{-1}$, $\textstyle{k_7=25.5}$ molecules$^{-1}$ hour$^{-1}$, $\textstyle{k_8=132.6}$ hour$^{-1}$ and $\textstyle{k_9=1}$ molecules hour$^{-1}$.)\\

\newpage 

\begin{center}
\includegraphics[width=6in]{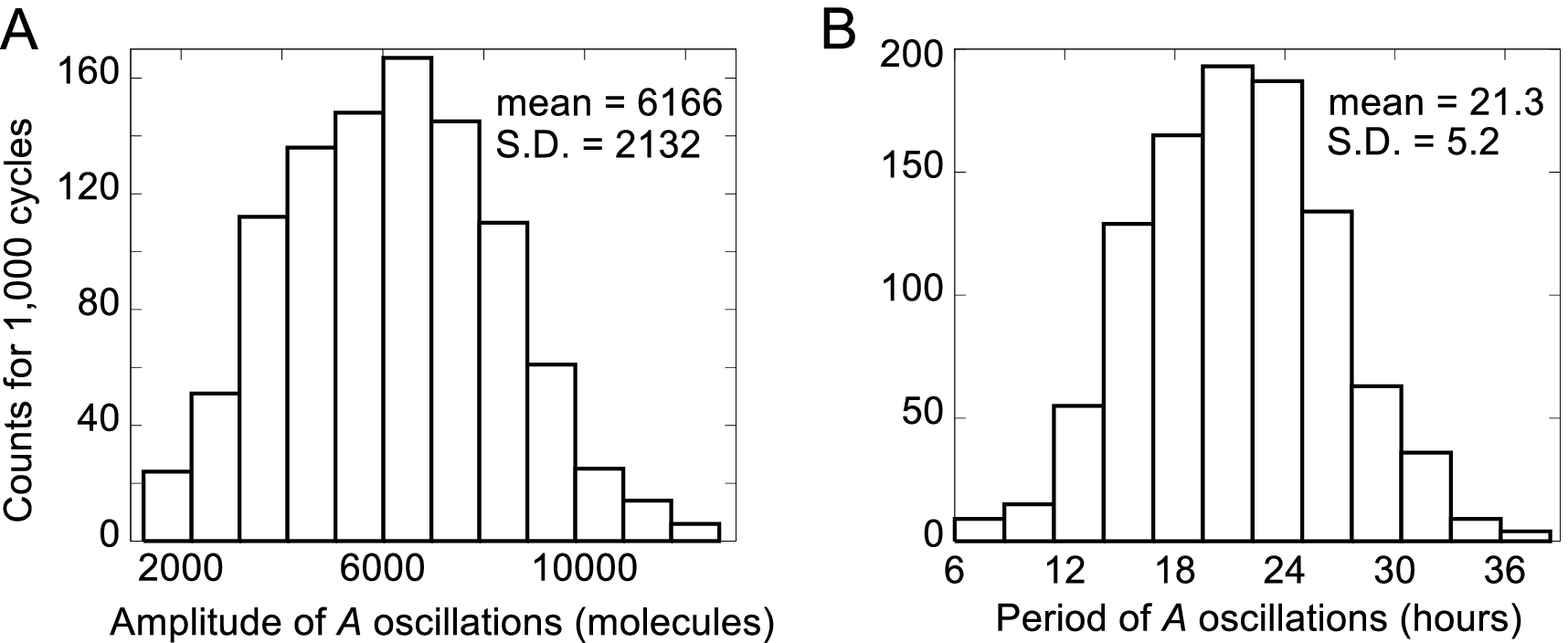}
\end{center}

\noindent {\bf Figure S2. Amplitude and period histograms of the stochastic simulation of $\boldsymbol{A}$.} {\bf A, B.} Amplitude and period histograms of the stochastic simulation of $A$, respectively. The values of the parameters are as in the section \textit{Methods: Biochemical reactions and rates} but now we set $\textstyle{k_4=1000}$ hour$^{-1}$, $\textstyle{k_5=5000}$ hour$^{-1}$, $\textstyle{k_7=25.5}$ molecules$^{-1}$ hour$^{-1}$, $\textstyle{k_8=132.6}$ hour$^{-1}$ and $\textstyle{k_9=1}$ molecules hour$^{-1}$. ({\bf A} and {\bf B} were calculated for 1,000 successive cycles. We assumed that a cycle occurs if the number of proteins $A$ increases to 1,000 molecules and then decreases to 700 molecules. The amplitude was calculated as the greatest number of $A$ molecules in each cycle. The period was calculated as the time interval that it takes the numbers of proteins $A$ to reach 1,000 molecules for the first time in two successive cycles.)\\

\newpage 

\begin{center}
\includegraphics[width=6in]{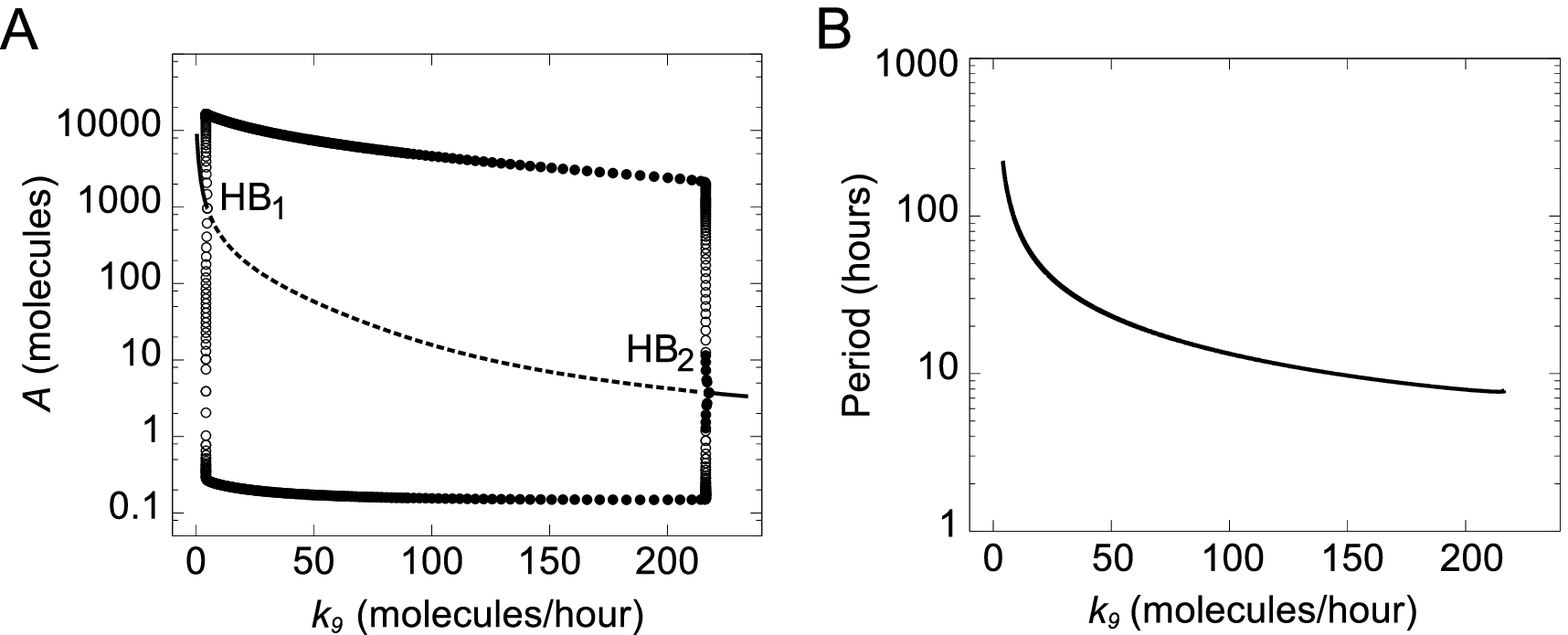}
\end{center}

\noindent {\bf Figure S3. Bifurcation diagram of the reduced model.} {\bf A.} Bifurcation diagram depending on $k_9$. The solid/dashed line represents stable/unstable fixed points. Black/white circles are the maximum and minimum values of $A$ during unstable/stable oscillations. HB denotes a Hopf Bifurcation point. HB$_1$ and HB$_2$ appear when the value of $k_9$ is 4.78 and 217.6 molecules hour$^{-1}$, respectively. {\bf B.} Period of the stable oscillations in {\bf A}.\\

\newpage 

\begin{center}
\includegraphics[width=3.5in]{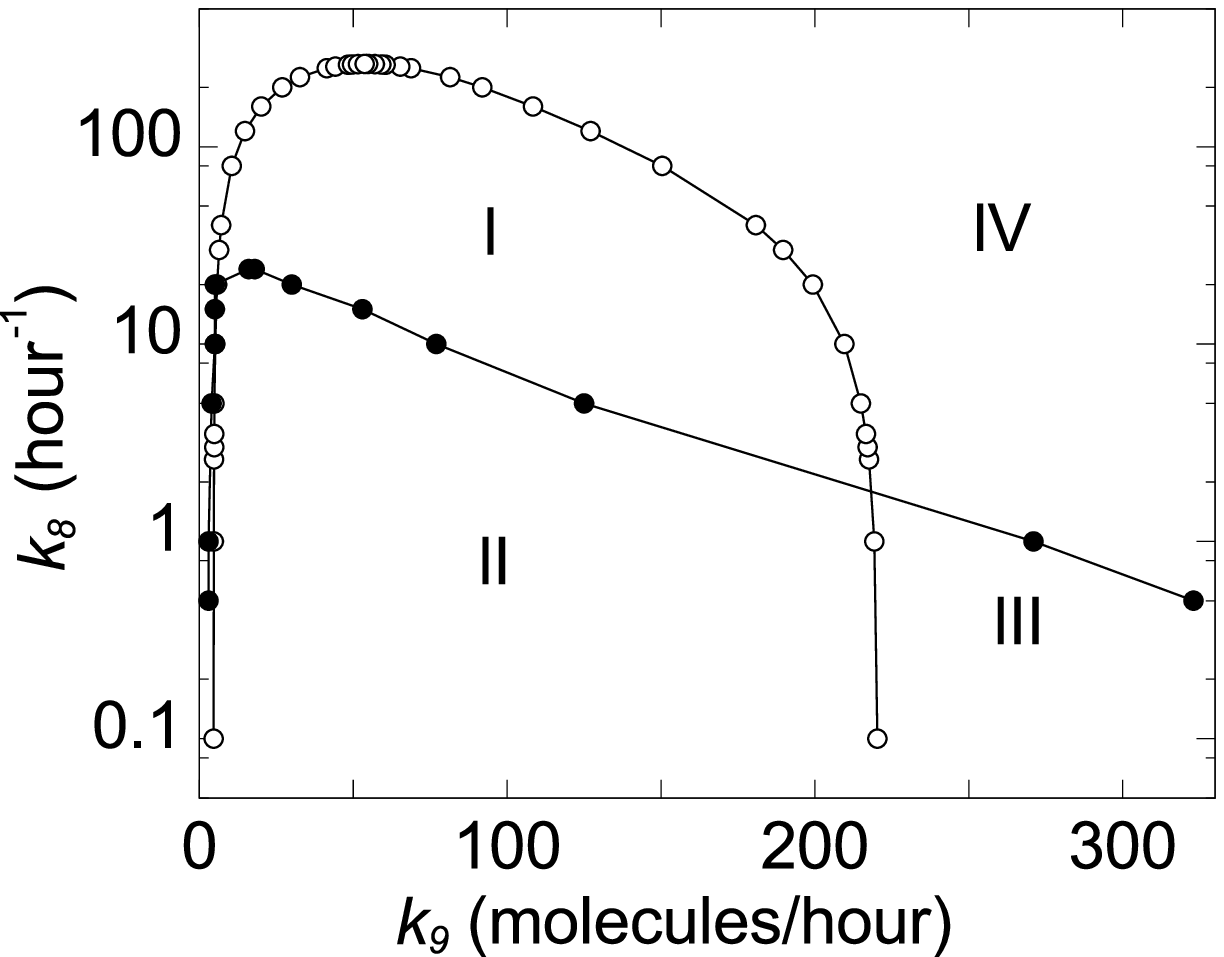}
\end{center}

\noindent {\bf Figure S4. Oscillatory regions in the reduced and stochastic models depending on $\boldsymbol{k_8}$ and $\boldsymbol{k_9}$.} {\bf Region I}. Oscillations in reduced model. {\bf Region II}. Oscillations in both reduced and stochastic model. {\bf Region III}. Oscillations in the stochastic model. {\bf Region IV}. No oscillations in any model. White circles represent the locus of Hopf bifurcations in the reduced model (data are presented in Table S2). Black circles represent locus of oscillations in the stochastic simulation (data are presented in Table S3). We assumed in the stochastic case that oscillations occur in a region if the correlation in the first period is greater than 0.2. (The lines connecting circles are designed to clearly single out the different regions.)\\

\newpage 

\begin{center}
\includegraphics[width=5in]{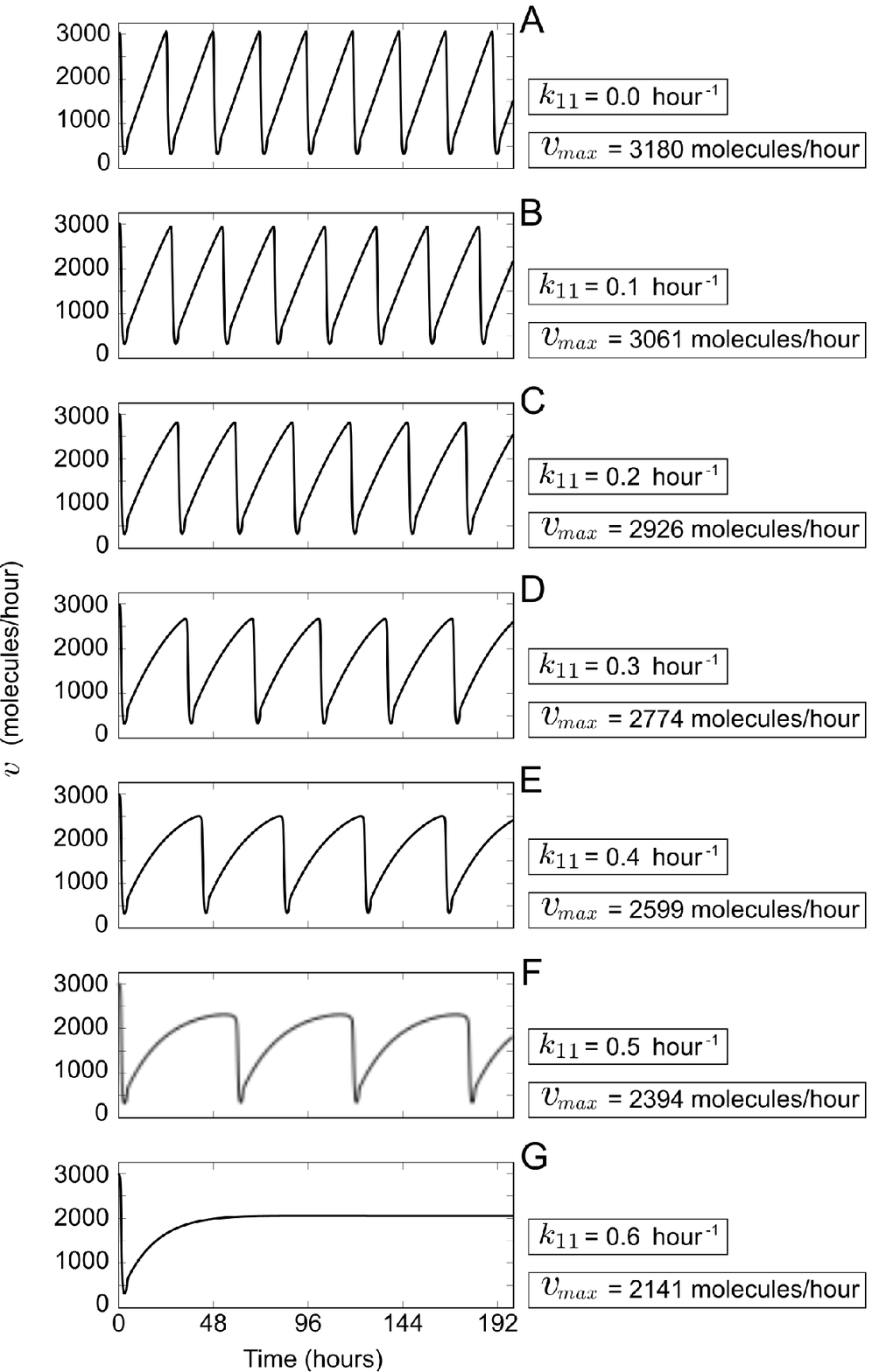}
\end{center}

\noindent {\bf Figure S5. Rate of the negative interaction for different values of $\boldsymbol{k_{11}}$.} Rate of the negative interaction ($v = k_8 C$) for different values of $k_{11}$, where $k_{11}$ is the rate of reaction $\textstyle{C \xrightarrow{} \phi}$. Deterministic simulations {\bf A, B, C, D, E, F} and {\bf G} correspond to $\textstyle{k_{11}}$ equals 0.0, 0.1, 0.2, 0.3, 0.4, 0.5 and 0.6 hour$^{-1}$, respectively. The values of the other parameters are as in the section \textit{Methods: Biochemical reactions and rates}. The oscillations stop when $\textstyle{k_{11}}=$ 0.6 hour$^{-1}$ ({\bf G}). If $\textstyle{k_{11}}$ is increased, $v$ increases slower, and its maximum value ($v_{max}$) is lower. The value of $v_{max}$ corresponds to the peak of the oscillations ($v_{max}$ is the value of the steady state in {\bf G}).\\

\newpage 

\begin{center}
\includegraphics[width=3.5in]{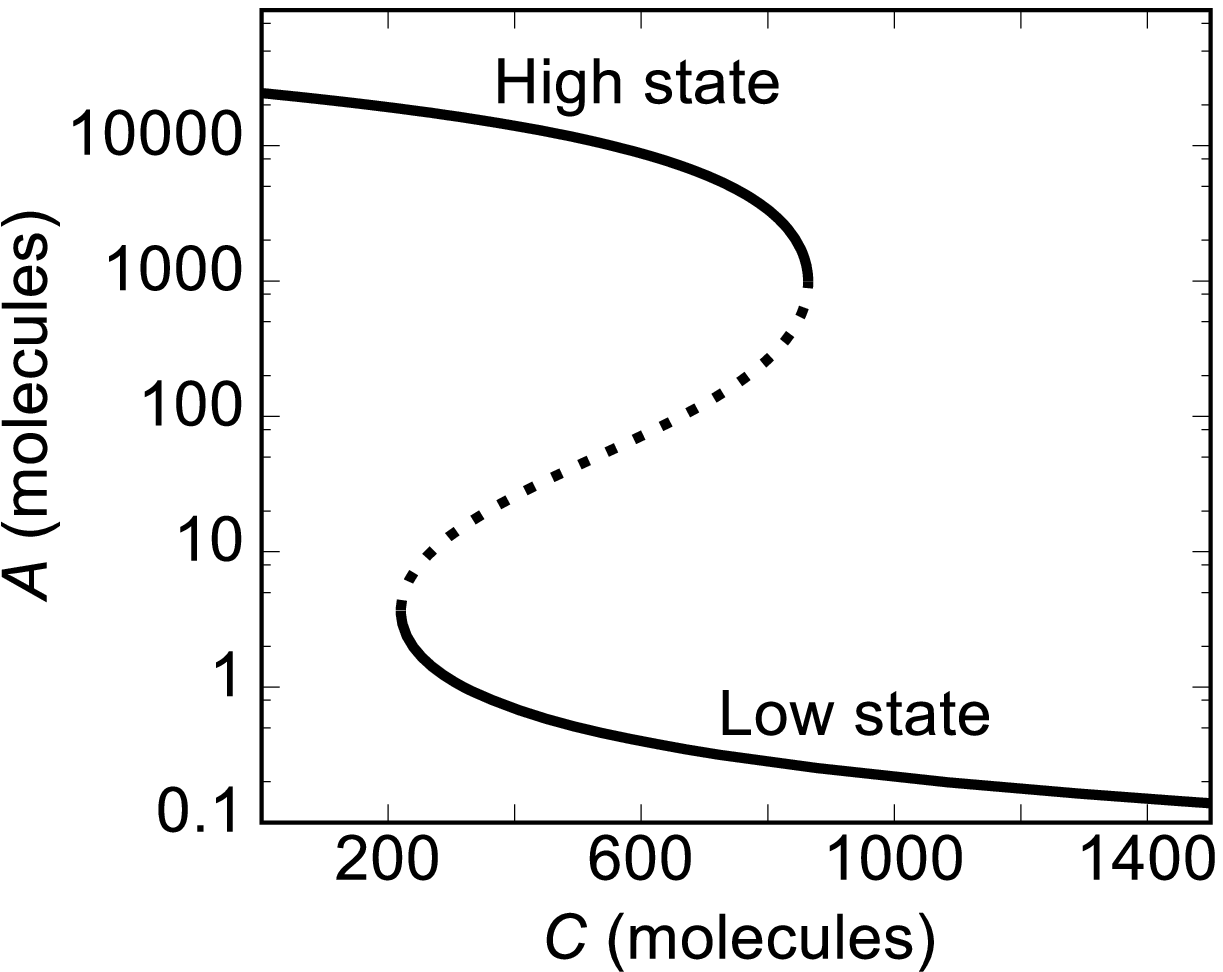}
\end{center}

\noindent {\bf Figure S6. Hysteresis diagram.} Hysteresis diagram depending on $C$. The curve is the solution of the equation $dA/dt=0$, where $C$ is assumed constant. The two solid lines in the diagram are the two stable steady states ``high'' and ``low'' as a function of $C$. The dashed line represents the unstable points in the diagram.\\

\newpage 

\section*{Supporting Information: Tables}

\noindent {\bf Table S1. Data points of Fig. 4I.}
\begin{table}[!ht]
\centering
\begin{tabular}{|c|c|c|c|c|}
\hline
& \multicolumn{2}{|c|}{\bf Period of $\boldsymbol{A}$} & \multicolumn{2}{|c|}{\bf Amplitude of $\boldsymbol{A}$}\\ 
\hline
\textbf{Changed rate} & \textbf{Mean (hours)} & \textbf{S.D. (hours)} & \textbf{Mean (molecules)} & \textbf{S.D. (molecules)}\\
\hline
none & 24.3 & 1.7 & 6723 & 858\\
\hline
$1.15k_1$ & 23.9 & 1.8 & 6554 & 909\\
\hline
$0.85k_1$ & 24.6 & 1.6 & 6792 & 867\\
\hline
$1.15k_{-1}$ & 24.5 & 1.6 & 6762 & 822\\
\hline
$0.85k_{-1}$ & 23.9 & 1.9 & 6528 & 974\\
\hline
$1.15k_2$ & 23.6 & 1.7 & 6365 & 879\\
\hline
$0.85k_2$ & 24.8 & 1.7 & 7017 & 865\\
\hline
$1.15k_3$ & 27.5 & 1.8 & 8796 & 1042\\
\hline
$0.85k_3$ & 21.0 & 1.7 & 4760 & 740\\
\hline
$1.15k_4$ & 21.9 & 1.6 & 5263 & 731\\
\hline
$0.85k_4$ & 27.2 & 1.8 & 8714 & 1082\\
\hline
$1.15k_5$ & 26.9 & 1.8 & 8458 & 1047\\
\hline
$0.85k_5$ & 21.6 & 1.7 & 5053 & 712\\
\hline
$1.15k_6$ & 23.6 & 1.6 & 6247 & 803 \\
\hline
$0.85k_6$ & 24.8 & 1.8 & 7138 & 1020\\
\hline
$1.15k_7$ & 24.6 & 1.7 & 6807 & 860\\
\hline
$0.85k_7$ & 23.9 & 1.8 & 6501 & 959\\
\hline
$1.15k_8$ & 21.5 & 1.6 & 5931 & 880\\
\hline
$0.85k_8$ & 27.8 & 1.8 & 7597 & 881\\
\hline
$1.15k_9$ & 21.9 & 1.6 & 6021 & 859\\
\hline
$0.85k_9$ & 27.3 & 1.8 & 7485 & 937\\
\hline
$1.15k_{10}$ & 24.3 & 1.6 & 6886 & 880\\
\hline
$0.85k_{10}$ & 24.3 & 1.7 & 6486 & 889\\
\hline
\end{tabular}
\begin{flushleft}
\end{flushleft}
\label{tab:label}
\end{table}

\newpage 

\noindent {\bf Table S2. Data points of locus Hopf bifurcation in reduced model (Fig. S4).}
\begin{table}[!ht]
\centering
\begin{tabular}{|c|c|c|}
\hline
$\boldsymbol{k_8}$ \textbf{(hour}$\boldsymbol{^{-1}}$\textbf{)} & $\boldsymbol{k_9^{min}}$ \textbf{(molecules hour}$\boldsymbol{^{-1}}$\textbf{)} & $\boldsymbol{k_9^{max}}$ \textbf{(molecules hour}$\boldsymbol{^{-1}}\textbf{)}$\\
\hline
0.1  &  4.65   &    220.3\\
\hline
1   &   4.70   &    219.3\\
\hline
2.6  &  4.78   &    217.6\\
\hline
3     & 4.80   &    217.1\\
\hline
3.5  &  4.82   &    216.6\\
\hline
5    &  4.90   &    214.9\\
\hline
10   &  5.17   &    209.6\\
\hline
20   &  5.75   &    199.3\\
\hline
30   &  6.39   &    189.7\\
\hline
40  &   7.10   &    180.8\\
\hline
80   &  10.5   &    150.4\\
\hline
120  &  14.8   &    127.1\\
\hline
160  &  20.1   &    108.4\\
\hline
200  &  26.9   &    91.9\\
\hline
225  &  32.7   &    81.5\\
\hline
250  &  41.4   &    68.8\\
\hline
255  &  44.2   &    65.3\\
\hline
260  &  48.3   &    60.5\\
\hline
261  &  49.6   &    59.1\\
\hline
262  &  51.5   &    57.0\\
\hline
262.6 & 53.7   &    54.8\\
\hline
\end{tabular}
\begin{flushleft}
\end{flushleft}
\label{tab:label}
\end{table}

\newpage 

\noindent {\bf Table S3. Data points of locus of oscillations with less than 20\% of correlation in the first period in the stochastic model (Fig. S4).}
\begin{table}[!ht]
\centering
\begin{tabular}{|c|c|c|}
\hline
$\boldsymbol{k_8}$ \textbf{(hour}$\boldsymbol{^{-1}}$\textbf{)} & $\boldsymbol{k_9^{min}}$ \textbf{(molecules hour}$\boldsymbol{^{-1}}$\textbf{)} & $\boldsymbol{k_9^{max}}$ \textbf{(molecules hour}$\boldsymbol{^{-1}}\textbf{)}$\\
\hline
0.5  &  3    &    323\\
\hline
1   &   3    &    271\\
\hline
5   &   4    &    125\\
\hline
10  &   5    &    77\\
\hline
15  &   5    &    53\\
\hline
20  &   5    &    30\\
\hline
24  &   16   &    18\\
\hline
\end{tabular}
\begin{flushleft}
\end{flushleft}
\label{tab:label}
\end{table}

\newpage 

\section*{Supporting Information: Text}

\noindent{\bf Text S1. The negative interaction does not produce an effective high nonlinearity.}\\

It has been demonstrated that protein sequestration produces an effective high nonlinearity \mbox{\cite{Buchler_a,Buchler_b}}. But this high nonlinearity is not observed if the repressor molecule is recycled (see equation S9 and figure S5 in \mbox{\cite{Buchler_a}}). The biochemical reactions that describe the negative interaction are as follows: 

\begin{equation}
\label{reactions_NI}
\begin{alignedat}{2}
&\text{$A$ creation (or entry): } && \xrightarrow{f}  A \\
&\text{$A$ degradation: } && A \xrightarrow{k_6} \phi \\
&\text{Complex creation: } && R + A \xrightarrow{k_7} C \\
&\text{Complex decay into $R$: } && C \xrightarrow{k_8} R \\
&\text{$R$ creation (or entry): } && \phi \xrightarrow{k_9} R \\
&\text{$R$ degradation (or exit): } && R \xrightarrow{k_{10}} \phi.
\end{alignedat}
\end{equation}

The dynamics of these reactions are described by the following EDOs:

\begin{equation}
\label{edos_reduced_NI}
\begin{aligned}
dA/dt & = f - k_6 A - k_7 A R\\
dR/dt & = - k_7 A R + k_8 C + k_9 - k_{10} R\\
dC/dt & = k_7 A R - k_8 C,
\end{aligned}
\end{equation}

As in \mbox{\cite{Buchler_a}}, these equations can be solved at steady state to yield:

\begin{equation}
\label{edos_reduced_NI_flux}
\begin{aligned}
A & = \frac{fk_{10}}{k_7k_9+k_6k_{10}} \\
R & = \frac{k_9}{k_{10}}\\
C & = \frac{fk_7k_9}{(k_7k_9+k_6k_{10})k_8},
\end{aligned}
\end{equation}
where we observe no nonlinearity in output $A$ as a function of input flux $f$.

\newpage 

\section*{Supporting Information: Files}

\noindent {\bf File S1. Complete deterministic model (XPPAUT software).}\\

\noindent Available in:\\
\url{http://www.plosone.org/article/fetchSingleRepresentation.action?uri=info:doi/10.1371/journal.pone.0027414.s011}\\

\noindent {\bf File S2. Reduced deterministic model (XPPAUT software).}\\

\noindent Available in:\\
\url{http://www.plosone.org/article/fetchSingleRepresentation.action?uri=info:doi/10.1371/journal.pone.0027414.s012}\\

\noindent {\bf File S3. Stochastic and deterministic model (CAIN software).}\\

\noindent Available in:\\
\url{http://www.plosone.org/article/fetchSingleRepresentation.action?uri=info:doi/10.1371/journal.pone.0027414.s013}

\end{document}